\documentclass[twocolumn,twoside]{IEEEtran}
\usepackage{amsmath,amsfonts,amssymb,bm,mathrsfs}
\usepackage{graphicx,color}
\usepackage{soul} % Use \hl{} to yellow-highlight a portion of text
\usepackage[utf8]{inputenc}
\usepackage[T1]{fontenc}
\graphicspath{{Figs/}}
\definecolor{MidRed}{rgb}{0.78,0,0}
\definecolor{MidGreen}{rgb}{0,0.65,0}
\definecolor{MidBlue}{rgb}{0,0,0.68}

\newcommand{\unit}[1]{\ensuremath{\mathrm{#1}}}

\newcommand{\Celsius}{\ensuremath{\mathrm{^\circ C}}}
\newcommand{\req}[1]{(\ref{#1})}

\newcommand{\sbb}{\mathsf{b}}
\newcommand{\sh}{\mathsf{h}}
\newcommand{\sk}{\mathsf{k}}

\newcommand{\sx}{\mathsf{x}}
\newcommand{\sy}{\mathsf{y}}
\newcommand{\sz}{\mathsf{z}}

\newcommand{\xp}[1]{\mathbb{E}\left\{#1\right\}}

\renewcommand{\hl}[1]{#1}

\def\rboth{\sffamily\tiny } 
\begin{document}

\title{Frequency Stability Measurement of Cryogenic Sapphire Oscillators with a Multichannel Tracking DDS and the Two-Sample Covariance}

\author{Claudio E. Calosso$^\nabla$, Fran\c{c}ois Vernotte$^\exists$, Vincent Giordano$^\exists$,\\ 
Christophe Fluhr$^\otimes$, Beno\^{i}t Dubois$^\otimes$,  Enrico Rubiola$^{\exists\,\nabla\,\forall}$
\thanks{$^\nabla$ Physics Metrology Division, Istituto Nazionale di Ricerca Metrologica INRiM, Torino, Italy.}
\thanks{$\exists$ FEMTO-ST Institute, Dept.\ of Time and Frequency, Universit\'{e} de Bourgogne and Franche-Comt\'{e} (UBFC), and CNRS.  
Address: ENSMM, 26 Rue de l'Epitaphe, Besan\c{c}on, France.}
\thanks{$\otimes$ FEMTO Engineering, Besan\c{c}on, France.}
\thanks{$\forall$ ER is the reference author. 
E-mail: rubiola@femto-st.fr, home page http://rubiola.org.}}

\maketitle

%=======================================
\begin{abstract}
%=======================================
\boldmath
This article shows the first measurement of three 100 MHz signals exhibiting fluctuations from $2{\times}10^{-16}$ to parts in $10^{-15}$ for integration time $\tau$ between 1 s and 1 day. 
Such stable signals are provided by three Cryogenic Sapphire Oscillators (CSOs) operating at about 10 GHz, also delivering the 100 MHz output via a dedicated synthesizer.  The measurement is made possible by a 6-channel Tracking DDS (TDDS) and the two-sample covariance tool, used to estimate the Allan variance.  The use of two TDDS channels per CSO enables high rejection of the instrument background noise.  The covariance outperforms the Three-Cornered Hat (TCH) method in that the background converges to zero ``out of the box,'' with no need of the hypothesis that the instrument channels are equally noisy, nor of more sophisticated techniques to estimate the background noise of each channel.  Thanks to correlation and averaging, the instrument background (AVAR) rolls off with a slope $1/\sqrt{m}$, the number of \hl{measurements}, down to $10^{-18}$ at $\tau=10^4$ s.  
For consistency check, we compare the results to the traditional TCH method beating the 10 GHz outputs down to the \hl{MHz} region. 
Given the flexibility of the TDDS, our methods find immediate application to the measurement of the 250 MHz output of the FS combs.
\end{abstract}\unboldmath

%=========================================
\section{Introduction}
%=========================================
This article is made possible by the simultaneous availability in the same place of three Cryogenic Sapphire Oscillators (CSOs), a 6-channel Tracking DDS (TDDS) for the measurement of time fluctuations, and the know-how of clock statistics.

We demonstrate the first frequency-stability measurement of the three CSOs, taken simultaneously at the 10 GHz frequency of the oscillator loop and at the 100 MHz output of the dedicated synthesizer. 
The CSOs exhibit short-term fluctuations from $2{\times}10^{-16}$ to parts in $10^{-15}$ (Allan deviation, ADEV) for measurement time $\tau$ between 1 s and 1 day.  The synthesizer introduces a very small degradation to the purity of the microwave signal, and only for shortest $\tau$, one minute or less. 
Thus, the CSO exceeds by two orders of magnitude the short-term stability of H masers and other commercial atomic standards. 
The measurement of such CSOs is a challenging task because the target is significantly lower than the background noise of commercial instruments, and possible only by comparing three similar units.
For this reason, the measurement was until now done by beating the microwave outputs, with no synthesizer \cite{Hartnett-2010-UFFC,Fluhr-2016-UFFC}.  The use of ${\approx}$1--10 MHz beat notes relaxes the noise requirement for the instrument by 60--80 dB\@.  In this work, we demonstrate the direct ADEV measurement of the individual CSO at 100 MHz, with no need for the beat note method.  
  
A reliable and stable signal is of paramount importance in strategic facilities, where the short-term stability (up to a few hours) of the Hydrogen maser is not sufficient.  
For example, the accuracy of VLBI \cite{Nand-2011-MTT,Doeleman-2011-ASP,Bara-2012-MWCL} would be improved with the use of the CSO\@.  
The CSOs have been developed at JPL for the Cassini mission \cite{Dick-2000-UFFC}, and used by ESA in the space station in Malarg\"{u}e, Argentina \cite{Giordano-2012-RSI}.
The CSO proved to be the best flywheel for the Cesium fountains in a time scale \cite{Heo-2016,Abgrall-2016-UFFC,Ikegami-2016-JPCS}, and the benefit for the ground stations of the Global Navigation Satellite System is just obvious.
A 100 MHz reference is easy to distribute with regular low-temperature-coefficient Heliax cables (1 ppm/\Celsius). \hl{Attenuation (2.7 dB/100\,m at 100 MHz for 1/2'' cables) limits the range to a few hundred meters}.  By contrast, microwave and optical signals are complex to distribute, and at the present time do not fit the general requirements for a continuously running facility.  
The femtosecond laser locked to a Fabry-Perot cavity can provide a VHF signal with stability in competition with the CSO\@.  However the laser technology will probably win in the long run thanks to the optical clocks, reliability is still far from the requirements mentioned.   

The TDDS \cite{Calosso-2013-EFTF} is a radically new concept in frequency metrology.  In short, six DDSs are each PLL-ed to one input signal, extracting the phase information from the phase-control word.  At once, this eliminates the complexity of the dual-mixer system \cite{Allan-1975-IFCS,Allan-1976-NBS,Brida-2002-RSI}, mitigates the thermal instability by using only wide band components and filtering numerically at the output, and enables the simultaneous measurement of the six inputs at quite different frequencies scattered in a wide range (presently, 5--400 MHz).  
Our TDDS \cite{Calligaris-2015} exhibits a background noise of $1.5{\times}10^{-14}/\tau$ (ADEV) per channel at 100 MHz.

The most common approach for phase measurement with digital methods starts from sampling and digitizing the input signal
\cite{Grove-2004-IFCS,Ecker-2014-PhD,Sherman-2016-RSI,Devoe-2018-RSI,Yu-2018-RSI}. The TDDS is superior to this approach because the phase noise of ADCs is higher than that of a  DDS \cite{Calosso-2012-IFCS,Cardenas-2017-RSI,Cardenas-2018-PhD} and because the digital signal processing is done at the speed of the phase fluctuations, instead of at the carrier frequency.

It is worth mentioning that the general literature on the measurement of the Allan variance, and on the comparison of multiple clocks, is surprisingly old.  The digital methods are recent because of the availability of fast ADCs, and still limited by the flicker noise of the ADCs, which is of the order of $-110$ \unit{dBrad^2} (power spectral density at 1 Hz) in the best cases \cite{Cardenas-2017-RSI,Cardenas-2018-PhD}.  This is 20--30 dB higher than a double-balanced mixer.  Given the very small number of labs that have the technology of the CSO, the measurement at the 100 MHz output is a rather new problem. 

We use the two-sample covariance \cite{Vernotte-2016-IFCS} to reject the TDDS noise averaging on a large number of measures.  The covariance is superior to the TCH in that the background noise converges to zero by theorem, thus there is no need for sophisticated analysis to estimate and compensate the background noise.  Finally, we compare the results to the traditional TCH with the beat note method.

In the following Sections we go through the three tools, CSO, TDDS and covariance, we describe the experiment, and we discuss the results.

%==================================================
\section{The Cryogenic Sapphire Oscillator}
%==================================================
The Cryogenic Sapphire Oscillator (CSO) is a long-term project started in Besançon 25 years ago.  A few laboratory prototypes have been built, demonstrating a stability (overlapped ADEV, drift removed) of parts in $10^{-16}$ floor, parts in $10^{-15}$ for $\tau$ up to 1 day, and a sufficient reliability for 1--2 years of unattended operation. 
The complete machine consists of the oscillator (in strict sense), the refrigerator, a dedicated synthesizer, and control equipment \cite{Giordano-2016-JPCS}.  See also \cite{Locke-2008-RSI} for a general review.
The CSO is based on the following ideas.

The sapphire (\unit{Al_2O_3}) monocrystal is an ideal material for dielectric resonators because of its low loss, and good mechanical and chemical  properties.  It is hard (9 Mohs, by definition), stiff and stable, and suitable to precision machining.  Our 10-GHz Whispering-Gallery-Mode (WGM) resonators achieve routinely $Q$ of the order of two billions at liquid-He temperature.  The value depends on the electromagnetic mode, on the crystal size, and on the specimen.

Certain modes exhibit a natural turning point of the resonant frequency at 5--8 K, just above the He boiling point \cite{Jones-1988}.  This is due to the presence of paramagnetic \unit{Cr^{3+}}, \unit{Fe^{3+}} and \unit{Mo^{3+}} impurities.  The growth process provides the right amount of such impurities, with very similar results over at least two growth technologies and manufacturers \cite{Giordano-2016-MTT}.  The \unit{WGH_{15,0,0}} mode of a cylinder of 54 mm diameter and 30 mm height resonates close to 10 GHz.  \hl{The electromagnetic energy is confined in ${\approx}1/10$ of the volume in the outer perimeter, indeed in a volume large enough to keep the resonator in highly linear regime, and to provide high stability.} 

The resonator is cooled by a two-stage pulse-tube refrigerator \cite{Wang-1997}, and temperature stabilized by heating to the turning point within 100 $\mu$K.  The refrigerator is a special design exhibiting low vibes (${<1}~\mu\text{m}_\text{pp}$).

\begin{figure}[t]\centering
\includegraphics[width=70mm]{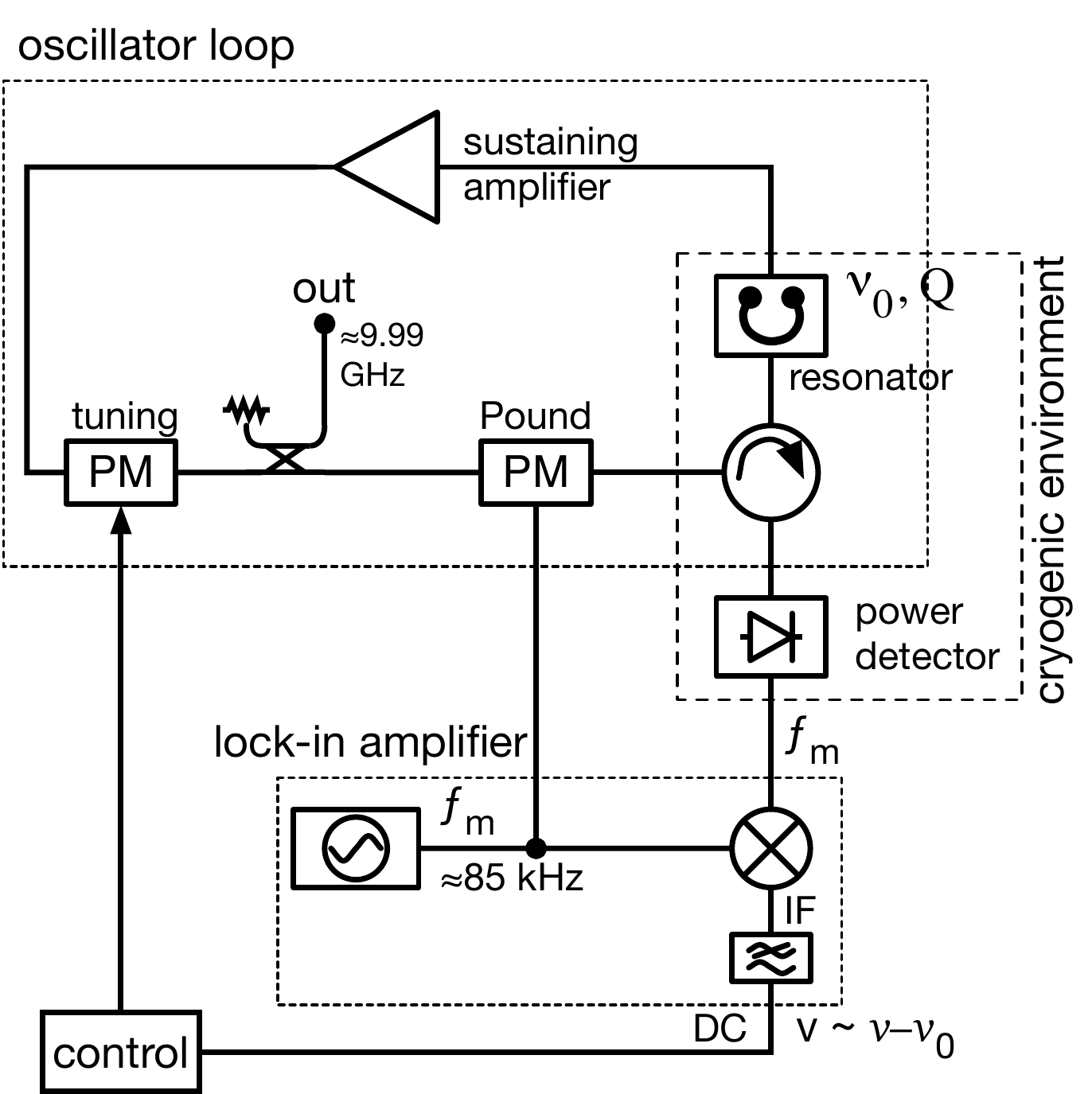}
\caption{Block diagram of the Pound-Galani cryogenic oscillator.}
\label{fig:Pound-Galani}
\end{figure}

The oscillator is a Pound-Galani, shown on Fig.~\ref{fig:Pound-Galani}.  
The Pound scheme \cite{Pound-1946} makes use of an auxiliary oscillator frequency-stabilzed to the main resonator.  The resonator is used in reflection mode, with phase modulation sidebands out of the resonator bandwidth.  The main vertues are the reduction of flicker and drift thanks to the AC modulation, and the inherent rejection of the fluctuations in the electrical path from the resonator to the detector.  The Galani version \cite{Galani-1984-MTT} implements the auxiliary oscillator using the same resonator, in transmission mode.  This solution provides higher $Q$ and better resonator stability, as compared to an external VCO\@.  Additionally, the Galani version is suitable to simple and effective design of the frequency control thanks to a pole-zero cancellation in the loop function, which results from using the same resonator in both oscillator and control.
A power control, not shown on Fig.~\ref{fig:Pound-Galani}, keeps the power constant within 3 ppm, preventing the fluctuations from degrading the stability via radiation pressure \cite{Chang-1997-PRL} and self heating.

\begin{figure}[t]\centering
\includegraphics[width=86mm]{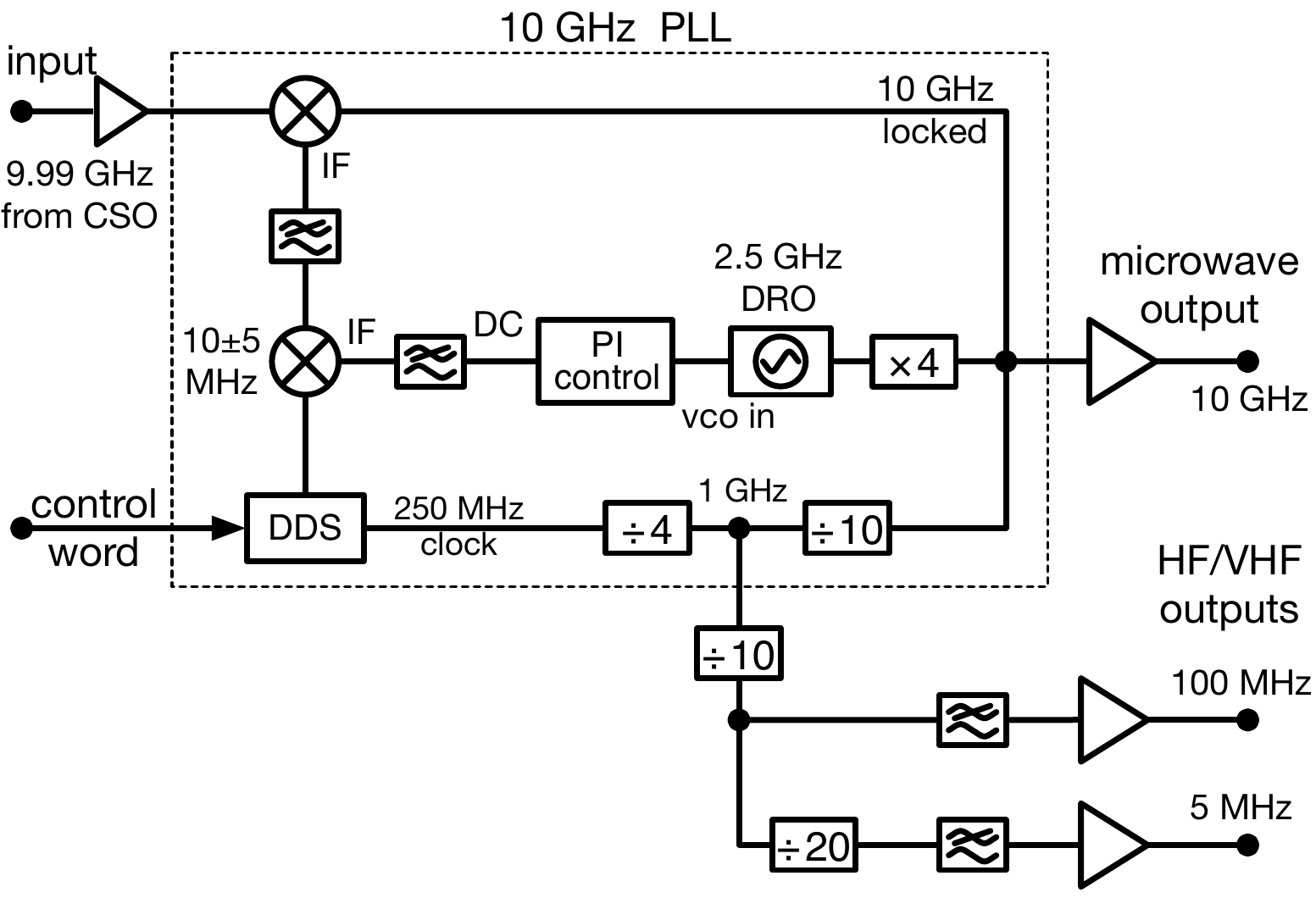}
\caption{Block diagram of the dedicated synthesizer.}
\label{fig:Synthesizer}
\end{figure}

\hl{The choice of the oscillation frequency is a key point of the design.  Modeling and machining the sapphire limits the initial accuracy to ${\approx}1$~MHz, and we prefer not to go through measurement-and-trimming iterations to achieve a more accurate value.
Such tolerances are useful in that they prevent electromagnetic interference between oscillators.
So, we machine the sapphire for $\nu_0=10~\unit{GHz} - \Delta$, where $\Delta=10{\pm}5$ MHz, and the synthesizer adds $\Delta$.  However, frequencies close to 5/10/15 MHz must be avoided for interference immunity.  It is wise to keep a margin of at least 10 kHz.}
The design turns out to be quite simple (Fig.~\ref{fig:Synthesizer}), yet achieving $\mu$Hz resolution at 10 GHz, and $10^{-4}$ tuning range. 
\hl{The specs for the time fluctuation of the DDS are relaxed by the ratio $\nu_0/\Delta\approx10^3$.  Thanks to this leverage factor, the DDS contributes ${\approx}10^{-17}$ to the stability.}

%==================================================
\section{The Multi-Channel Tracking DDS}
%==================================================
\begin{figure}\centering
\includegraphics[width=64mm]{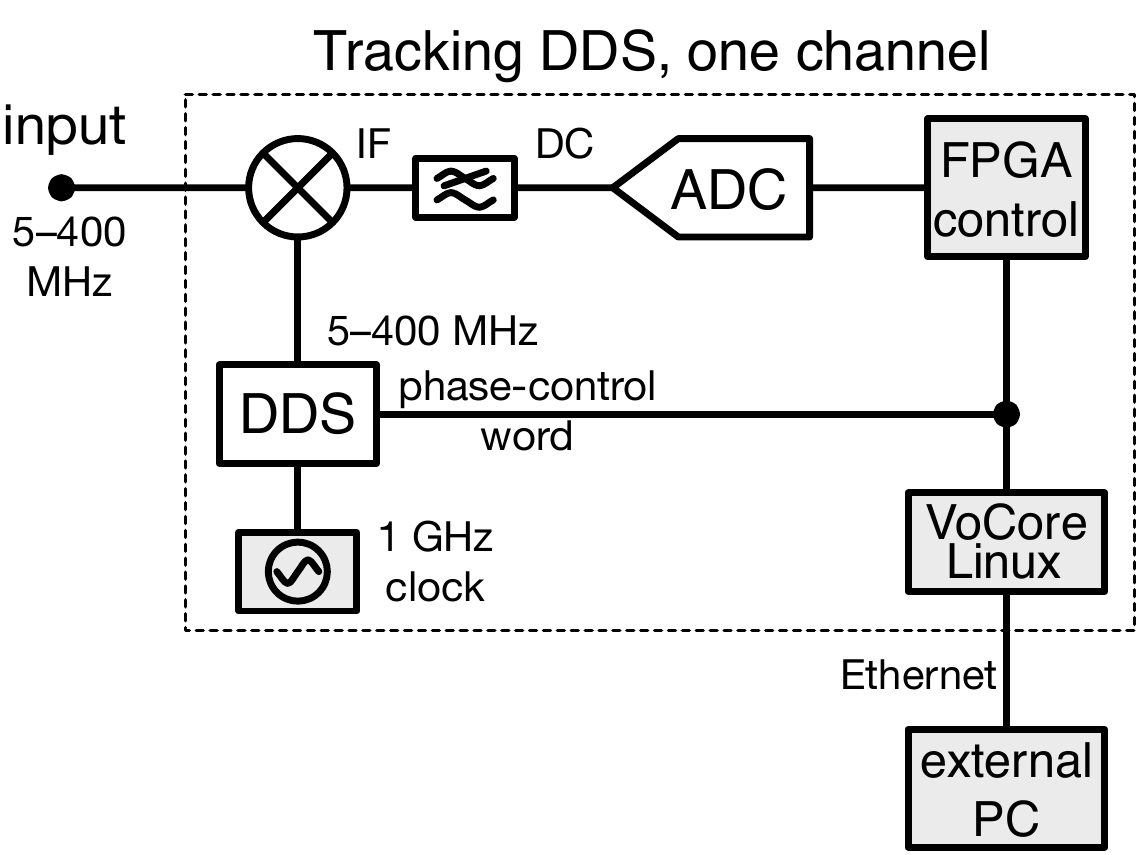}
\caption{Block diagram of the Tracking DDS.  The complete machine consists of 6 equal channels.  The grey blocks are shared by the 6 channels.}
\label{fig:TDDS-Scheme}
\end{figure}

\begin{figure}\centering
\includegraphics[width=88mm]{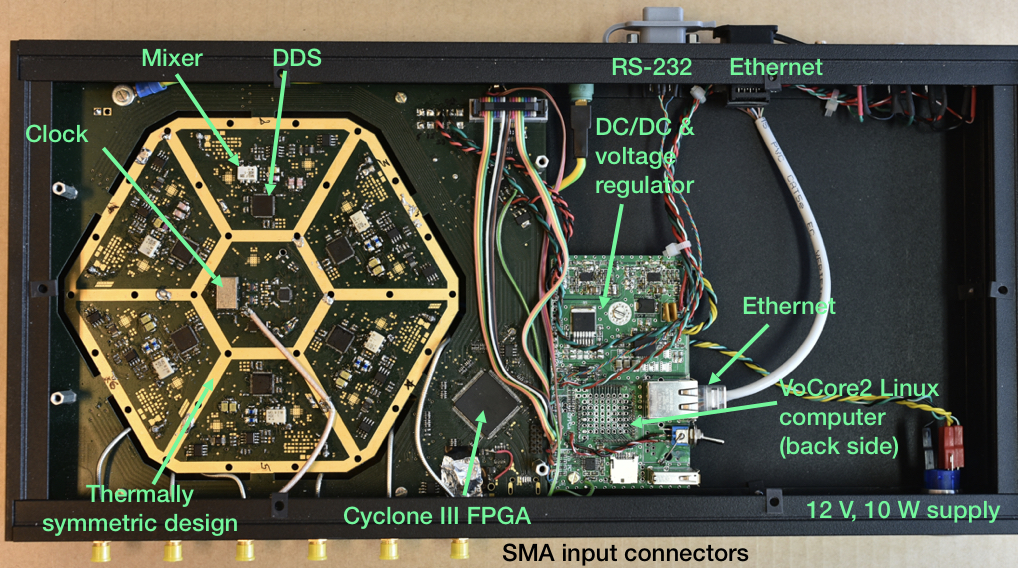}
\caption{Photo of the instrument.}
\label{fig:TDDS-Photo}
\end{figure}

Our instrument is a multi-channel real-time phasemeter based on the TDDS technique \cite{Calosso-2013-EFTF,Calligaris-2015}.  Figure~\ref{fig:TDDS-Scheme} shows the scheme of one channel, and Fig.~\ref{fig:TDDS-Photo} shows the complete machine.
A Proportional-Integral control implemented in FPGA phase-locks the DDS to the input acting on the phase-control word.  
Controlling the phase, instead of the frequency, requires a phase accumulator that counts the multiple cycles.
The discriminator is a double-balanced mixer Mini Circuits ADE-1.  
The acquisition and lock range is 5\ldots400 MHz.  The lower limit is set by the mixer, and the upper limit by the sampling frequency of the DDS (1 GHz).

The phase error is digitized on 16 bits at 500 kS/s.  This is also the sampling frequency of the digital control.  
The loop bandwidth is of 2--20 kHz, depending on internal parameters.  Anyway, the value is not critical.
\hl{Within the feedback-loop bandwidth, the phase-control word is equal to the phase difference between the input and the local clock.}  
The FPGA guarantees that all the measures are simultaneous.  
This is necessary to cancel the fluctuation of the internal 1 GHz clock, common to all the channels.  

The phase error is converted into a stream of phase-time data $\sx(t)$, low-passed at the cutoff frequency $f_H=5$ Hz, sampled at 10 S/s and transferred to the external PC by a VoCore2 Linux computer \cite{VoCore2}.

\begin{figure}\centering
\includegraphics[width=72mm]{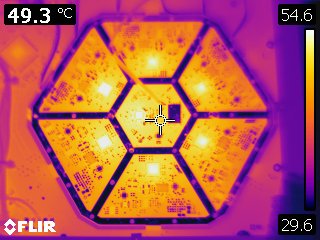}
\caption{Thermal image of the tracking DDS\@.}
\label{fig:Thermal-image}
\end{figure}

\begin{figure}\centering
\includegraphics[width=82mm]{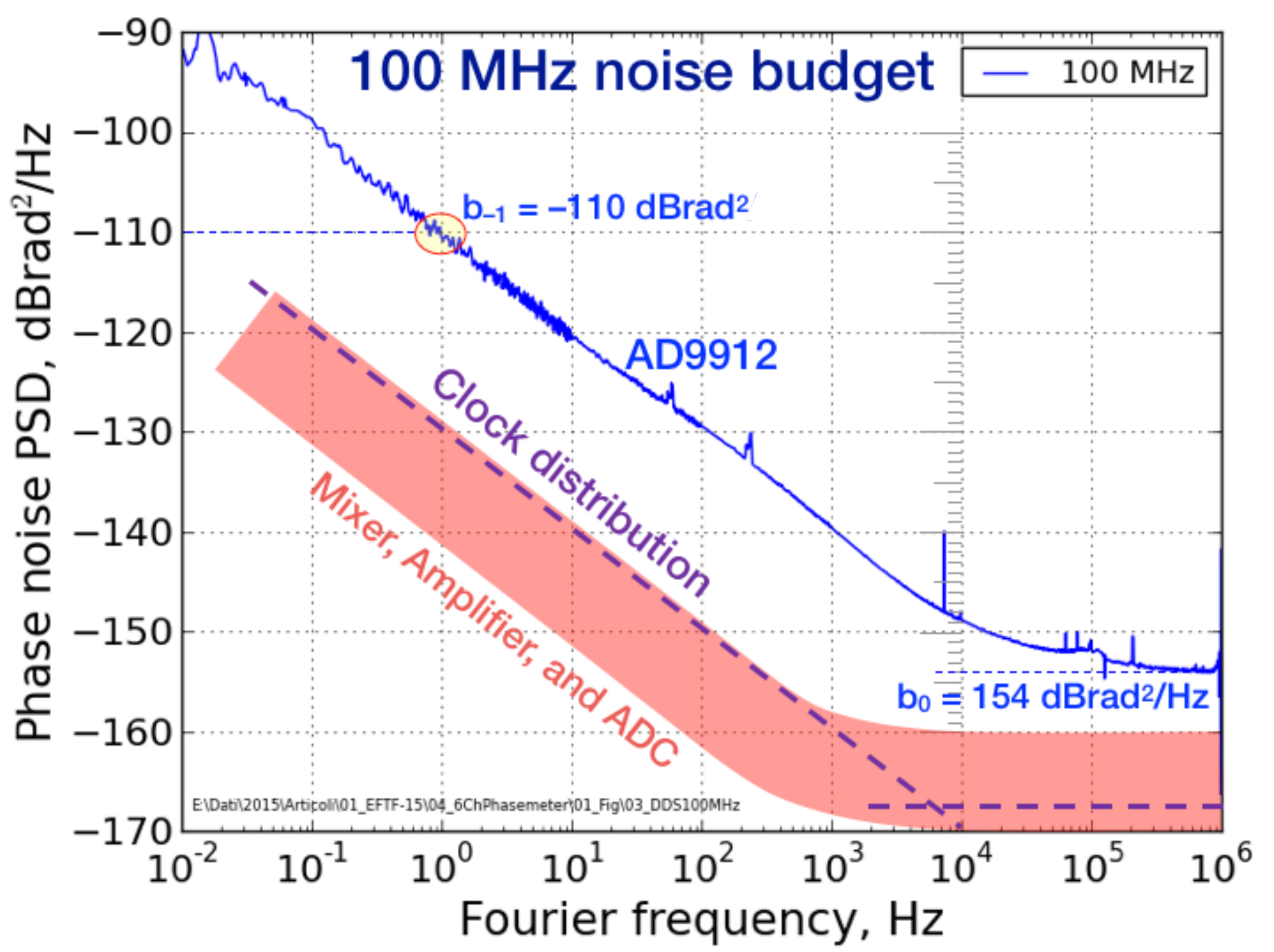}
\caption{Noise budget of the Tracking DDS at 100 MHz.}
\label{fig:TDDS-Noise}
\end{figure}

For best stability, we care about low dissipated power and geometrical symmetry.  The DDS is an Analog Devices AD9912, which has a typical dissipation of 650 mW, smaller than that of other high-frequency DDSs.  The FPGA is an Altera CYCLONE III with 25000 logic elements, 30\% of which are actually used.  For low dissipation, the 125 MHz clock is down converted to 10 MHz using an internal PLL\@.
The VoCore2 dissipates 1 W\@.
The complete instrument takes 10 W power from a single +12 V supply.
Figure~\ref{fig:Thermal-image} shows the thermal image of the six channels.  The temperature sensitivity is of about 1 ps/K (B-type uncertainty) on each channel, limited by the mixer.
  
Figure~\ref{fig:TDDS-Noise} shows the noise budget of the Tracking DDS at \hl{$\nu_\text{DDS}=100$} MHz carrier.  At low Fourier frequency, the $1/f$ PM noise of the DDS ($\sbb_{-1}=-110$ dBrad$^2$) dominates, being 20 dB higher than the noise of the clock distribution, and ${\approx}30$ dB higher than the noise of the mixer and the amplifier.  The DDS flicker noise is of the time type.  The $1/f$ term of $S_\sx(f)$ is \hl{$\sk_{-1}=\sbb_{-1}/(2\pi\nu_\text{DDS})^2=2.5{\times}10^{-29}~\mathrm{s}^2$}, independent of \hl{$\nu_\text{DDS}$}, while $\sbb_{-1}$ scales proportionally to \hl{$\nu_\text{DDS}$}.  Conversely, the $1/f$ noise of mixer and amplifier is of the phase type, with $\sbb_{-1}$ independent of \hl{$\nu_\text{DDS}$} in a wide range, and \hl{$\sk_{-1}\propto1/\nu_\text{DDS}$}.
The DDS $1/f$ noise is dominant from \hl{$\nu_\text{DDS}=10$} MHz to the maximum \hl{carrier} frequency.  At \hl{$\nu_\text{DDS}<10$} MHz, the $1/f$ noise of the mixer and of the amplifier is no longer negligible, and the background noise starts degrading. 
The quantity $\smash{\sqrt{\sk_{-1}}}=5$ fs is the flicker fluctuation of the DDS\@.
Converting the phase noise into frequency stability, with $f_H=5$ Hz we get $\sigma=1.5{\times}10^{-14}$ at $\tau=1$ s and 100 MHz carrier, with slope close to $1/\tau$.  This is the background noise of the instrument, one channel.

%==================================================
\section{Statistics}
\begin{figure}\centering
\includegraphics[width=74mm]{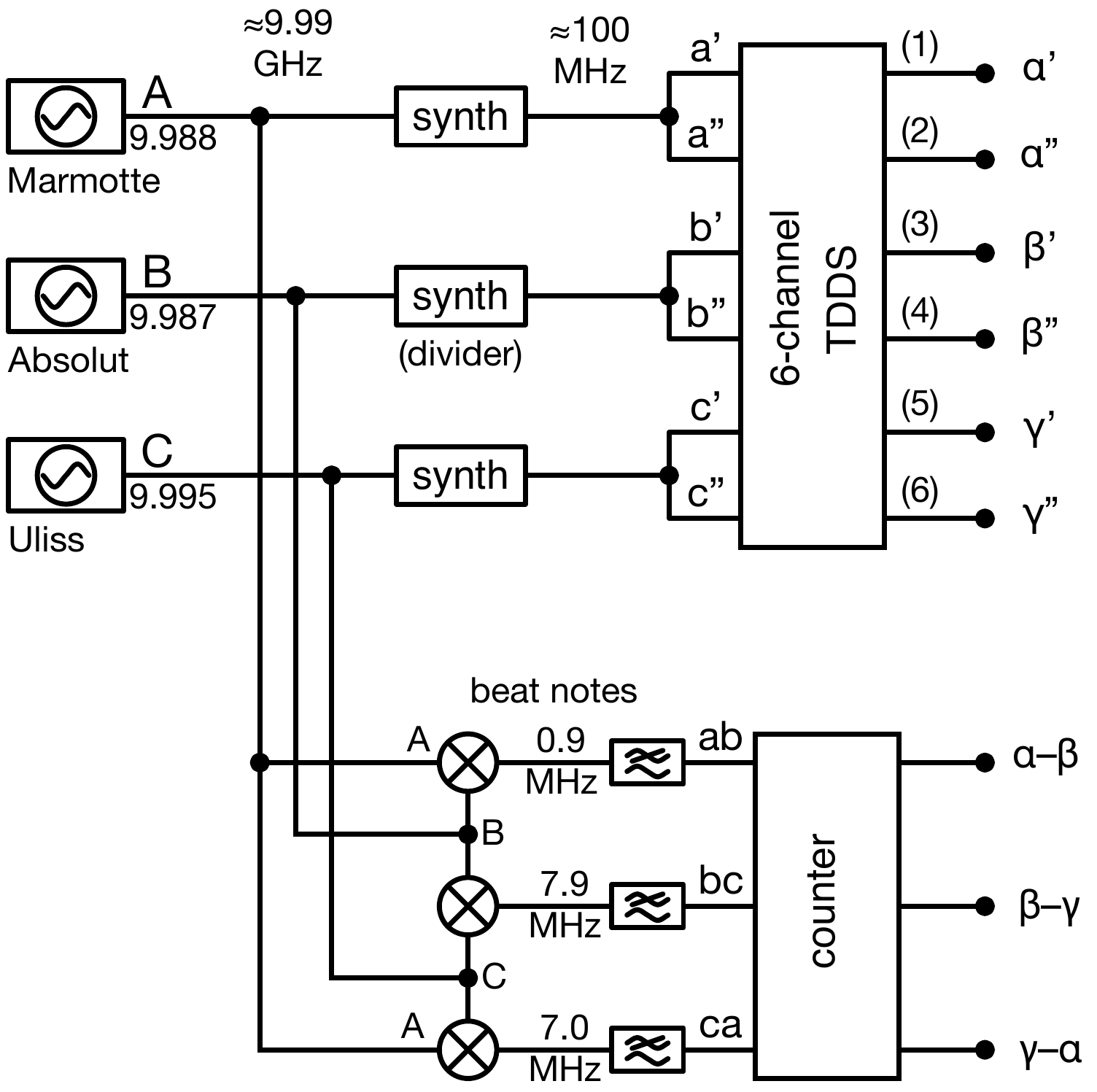}
\caption{Block diagram of the experiment.}
\label{fig:Experiment-scheme}
\end{figure}

%==================================================
The Allan variance can be  written as
\begin{align}
\sigma^2_\sy(\tau) = 
\xp{\frac{\left(\overline{\sy}_2-\overline{\sy}_1\right)^2}{2}}
=\xp{\frac{\left(\sx_2-2\sx_1+\sx_0\right)^2}{2\tau}}\,,
\label{eqn:avar}
\end{align}
where $\xp{\,}$ is the mathematical expectation, $\overline{\sy}$ is the fractional frequency fluctuation averaged over the measurement time $\tau$, and the subscripts `1' and `2' refer to contiguous time slots.
The alternate formulation in terms of the phase time $\sx$ sampled at regular intervals $\tau$ relates to the `second difference' method \cite{Greenhall-1989-UFFC}, equivalent to the method used here.  Making $\sx_2-2\sx_1+\sx_0$ explicit emphasizes that there is no dead time in $\overline{\sy}_2-\overline{\sy}_1$.  
Since all the variances in this article are $\sigma^2_\sy(\tau)$, we will omit $\tau$ and the subscript $\sy$, with no ambiguity.  We will have numerous occurrences of $\overline{\sy}_2-\overline{\sy}_1$ in the same formula, referring to different oscillators and instruments.  It is therefore appropriate to define the \emph{fractional frequency difference}
\begin{align}
\sz=\overline{\sy}_2-\overline{\sy}_1=
\frac{1}{\tau} \Big(\sx_2-2\sx_1+\sx_0\Big)\,.
\label{eqn:define-z}
\end{align}
Accordingly, the Allan variance is
\begin{align}
\sigma^2&=\frac{1}{2}\xp{\sz^2}\,.
\end{align}
The obvious extension to the two-sample covariance is
\begin{align}
\sigma_{ij}&=\frac{1}{2}\xp{\sz_i\sz_j}\,.
\end{align}
We denote the oscillators with the subscripts $A$, $B$ and $C$ (Capital); the instrument inputs with $a$, $b$ and $c$ (lowercase); and the instrument readouts with $\alpha$, $\beta$ and $\gamma$ (Greek), as seen on Fig.~\ref{fig:Experiment-scheme}.  
Following the path `A,' the oscillator's $\sz_A$ is sent to the instrument, which contributes $\sz_a$, and delivers the readout 
\begin{align}
\sz_\alpha=\sz_A+\sz_a\,.
\end{align}
This is similar to the Shannon channel, where the received signal is equal to the transmitted signal plus noise.  

All our statistical measurements rely on the hypothesis that all oscillators and instrument channels have statistically independent noise  processes, so that $\xp{\sz_i\sz_j}=0$ for $i\neq j$. Accordingly, 
\begin{align}
\sigma^2_\alpha=\sigma^2_A+\sigma^2_a\,,
\end{align}
and likewise for the other channels.

\subsection{Three-Cornered Hat Method}
%--------------------------------------
The TCH is a well established method to measure the variance of each oscillator by comparing three units \cite{Gray-1974}. 
The method is better understood by provisionally admitting that the counter noise is low enough to enable the direct measurement.
Because the internal reference of the instrument does not have sufficient stability for absolute measurements, we rely on the differences $\smash{\sx_{\beta-\alpha}=\sx_\beta-\sx_\alpha}$, $\smash{\sx_{\gamma-\beta}=\sx_\gamma-\sx_\beta}$ and $\smash{\sx_{\alpha-\gamma}=\sx_\alpha-\sx_\gamma}$ taken simultaneously. 
Using the corresponding fractional frequency differences, we calculate the variances
\begin{align}
\label{eqn:tch1}
\sigma^2_{\beta-\alpha}&=\frac{1}{2}\xp{\big(\sz_\beta-\sz_\alpha\big)^2}
=\sigma^2_B+\sigma^2_A+\sigma^2_b+\sigma^2_a\\[1ex]
\label{eqn:tch2}
\sigma^2_{\gamma-\beta}&=\frac{1}{2}\xp{\big(\sz_\gamma-\sz_\beta\big)^2}
=\sigma^2_C+\sigma^2_B+\sigma^2_c+\sigma^2_b\\[1ex]
\label{eqn:tch3}
\sigma^2_{\alpha-\gamma}&=\frac{1}{2}\xp{\big(\sz_\alpha-\sz_\gamma\big)^2}
=\sigma^2_A+\sigma^2_C+\sigma^2_a+\sigma^2_c\,.
\end{align}
\def\SystemTCH{
\begin{align}
\label{eqn:tch}
\left\{
\begin{array}{c}
\sigma^2_{\beta-\alpha}=\frac{1}{2}\xp{\big(\sz_\beta-\sz_\alpha\big)^2}
=\sigma^2_B+\sigma^2_A+\sigma^2_b+\sigma^2_a\\[1ex]
\sigma^2_{\gamma-\beta}=\frac{1}{2}\xp{\big(\sz_\gamma-\sz_\beta\big)^2}
=\sigma^2_C+\sigma^2_B+\sigma^2_c+\sigma^2_b\\[1ex]
\sigma^2_{\alpha-\gamma}=\frac{1}{2}\xp{\big(\sz_\alpha-\sz_\gamma\big)^2}
=\sigma^2_A+\sigma^2_C+\sigma^2_a+\sigma^2_c\,.
\end{array}\right.
\end{align}}
This is immediately seen by expanding the terms inside the $\xp{~}$ operator.  In \req{eqn:tch1}, $(\sz_\beta-\sz_\alpha)^2=(\sz_B+\sz_b-\sz_A-\sz_a)^2$.  The four `$\sz$'-s are statistically independent (separate oscillators and separate channels of the instrument), thus $\xp{\sz_i\sz_j}=0$ for all the cross terms $2\sz_i\sz_j$, $i\ne j$.  Likewise, \req{eqn:tch2} and \req{eqn:tch3}. 

Notice that the subscripts form a group $A{\rightarrow}B{\rightarrow}C{\rightarrow}A\ldots$, thus 
we can derive all equations for $B$ from the \hl{homologous} equation for $A$ by replacing $A{\rightarrow}B$, $a{\rightarrow}b$ and $\alpha{\rightarrow}\beta$ in the same equation for $A$.  Likewise for $C$ from $B$, and for $A$ from $C$.  This is clearly seen on \req{eqn:tch1}, \req{eqn:tch2} and \req{eqn:tch3}.  Hereafter, we will write only the equations for the oscillator $A$, because the other two equations can be written with the above rule. 

Solving \req{eqn:tch1}-\req{eqn:tch3}, the variance of the the oscillator $A$ is
\begin{align}
\sigma^2_A&=\frac{1}{2}\Big(\sigma^2_{\beta-\alpha}-\sigma^2_{\gamma-\beta}+\sigma^2_{\alpha-\gamma}\Big)-\frac{1}{2}\sigma^2_a\,.
%\\[1ex]
%\sigma^2_B&=\frac{1}{2}\left(\sigma^2_{\gamma-\beta}-\sigma^2_{\alpha-\gamma}+\sigma^2_{\beta-\alpha}\right)-\frac{1}{2}\sigma^2_b\\[1ex]
%\sigma^2_C&=\frac{1}{2}\left(\sigma^2_{\alpha-\gamma}-\sigma^2_{\beta-\alpha}+\sigma^2_{\gamma-\beta}\right)-\frac{1}{2}\sigma^2_b
\end{align}

The terms $\sigma^2_a$, $\sigma^2_b$ and $\sigma^2_c$ \hl{in {\req{eqn:tch1}}, {\req{eqn:tch2}} and {\req{eqn:tch3}} are the white and flicker PM noise introduced by the instrument}, thus they roll off as $1/\tau^2$.  
Nonetheless, they are annoying because no instrument has noise low enough to measure the short-term stability of our cryogenic oscillators.  For this reason, we have to combine the TCH with the beat method (the lower part of Fig.~\ref{fig:Experiment-scheme}).  Beating the $\nu_0$ output (${\approx}10$ GHz) down to \hl{the MHz region} relaxes the stability requirement by a factor $\nu_\text{beat}/\nu_0$, which is $10^{-4}\ldots10^{-3}$.

\subsection{Covariance Method}
%--------------------------------------
The two-sample covariance method is a different way to process of the same data, $\sx_{\beta-\alpha}$, $\sx_{\gamma-\beta}$ and $\sx_{\alpha-\gamma}$, measured simultaneously \cite{Fest-1983-IM,Vernotte-2016-IFCS}.
In this case, we exploit the product of fractional frequency differences 
\begin{align}
\label{eqn:cov-A}
\frac{1}{2}\xp{\big(\sz_\beta-\sz_\alpha\big)\big(\sz_\gamma-\sz_\alpha\big)}=\frac{1}{2}\xp{\sz_\alpha^2}=\sigma^2_A+\sigma^2_a
%\\[1ex]
%\label{eqn:cov-B}
%\frac{1}{2}\xp{\Big[\sz_\gamma-\sz_\beta\Big]\Big[\sz_\alpha-\sz_\beta\Big]}=\frac{1}{2}\xp{\sz_\beta^2}=\sigma^2_B+\sigma^2_b\\[1ex]
%\label{eqn:cov-C}
%\frac{1}{2}\xp{\Big[\sz_\alpha-\sz_\gamma\Big]\Big[\sz_\beta-\sz_\gamma\Big]}=\frac{1}{2}\xp{\sz_\gamma^2}=\sigma^2_C+\sigma^2_c
\end{align}
\hl{for the oscillator $A$, and likewise for the oscillators $B$ and $C$}\@.
The proof is immediate.  \hl{Assuming that $\sz_\alpha$, $\sz_\beta$ and $\sz_\gamma$} are statistically independent (separate oscillators and instrument channels), all the cross terms in $\xp{(\ldots)(\ldots)}$ \hl{are equal to zero}, and only the square term remains.  
The result is the same of the TCH, and not suitable to our purposes for the same reasons.

The background noise can be rejected by using two channels per oscillator.  Let us start with $A$, which splits into `prime' and `second' channel.  Thus 
$\xp{(\sz_\beta-\sz_\alpha)(\sz_\gamma-\sz_\alpha)}$ of \req{eqn:cov-A} becomes $\xp{(\sz_\beta-\sz_{\alpha'})(\sz_\gamma-\sz_{\alpha''})}$.   \hl{Assuming that $\sz_{a'}$ and $\sz_{a''}$ are statistically independent, it holds that $\xp{\sz_{\alpha'}\sz_{\alpha''}}=\xp{\sz^2_A}$.  
This solves the problem.}

A more efficient use of the hardware is possible.  Replacing $\sz_\beta\rightarrow(\sz_{\beta'}+\sz_{\beta''})/2$ and $\sz_\gamma\rightarrow(\sz_{\gamma'}+\sz_{\gamma''})/2$ results in lower background noise, hence in faster convergence
\begin{align}
\label{eqn:cov2-A}\!
\frac{1}{2}\xp{\left(\frac{\sz_{\beta'}+\sz_{\beta''}}{2}-\sz_{\alpha'}\right)\left(\frac{\sz_{\gamma'}+\sz_{\gamma''}}{2}-\sz_{\alpha''}\right)}
&=\sigma^2_A\,.
%\\[1em]
%\label{eqn:cov2-B}
%\frac{1}{2}\xp{\left[\frac{\sz_{\gamma'}+\sz_{\gamma''}}{2}-\sz_{\beta'}\right]\left[\frac{\sz_{\alpha'}+\sz_{\alpha''}}{2}-\sz_{\beta''}\right]}
%\nonumber\\=\frac{1}{2}\xp{\sz_B^2}
%&=\sigma^2_B\\[1em]
%\label{eqn:cov2-C}
%\frac{1}{2}\xp{\left[\frac{\sz_{\alpha'}+\sz_{\alpha''}}{2}-\sz_{\gamma'}\right]\left[\frac{\sz_{\beta'}+\sz_{\beta''}}{2}-\sz_{\gamma''}\right]}
%\nonumber\\=\frac{1}{2}\xp{\sz_C^2}
%&=\sigma^2_C
\end{align}
Averaging \req{eqn:cov2-A} with the same after interchanging $\alpha'$ with $\alpha''$ results in lower background noise.  Accordingly, the final equation we use is 
\begin{multline}
\label{eqn:cov2-A-final}
\frac{1}{4}\mathbb{E}\left\{
\left(\frac{\sz_{\beta'}+\sz_{\beta''}}{2}-\sz_{\alpha'}\right)\left(\frac{\sz_{\gamma'}+\sz_{\gamma''}}{2}-\sz_{\alpha''}\right)
\right.
+\\
~+\left.
\left(\frac{\sz_{\beta'}+\sz_{\beta''}}{2}-\sz_{\alpha''}\right)\left(\frac{\sz_{\gamma'}+\sz_{\gamma''}}{2}-\sz_{\alpha'}\right)\right\}
=\sigma^2_A\,.
\end{multline}

\subsection{Averaging on a Finite Data Record}
\hl{The mathematical expectation is replaced with the average on a finite time series of $m$ samples.}
We use the overlapped Allan variance in all cases.
The \hl{ultimate limit to the detection of fractional frequency fluctuations is}  
\begin{align}
\label{eqn:background-m}
\sigma&=\frac{\sigma_0}{\tau}\,\frac{1}{m^{1/4}}\,,
\end{align}
where $\sigma_0$ is the background at $\tau=1$ s, $m$ is the number of averages, and the ratio $\sigma_0/\tau$ is the usual `$1/\tau$' law for white and flicker PM noise\@.  
In turn, $m$ results from the duration $\mathcal{T}$ of the time series according to $m=\mathcal{T}/\tau$.  Combining the latter with \req{eqn:background-m}, we get 
\begin{align}
\label{eqn:background-T}
\sigma&=\frac{\sigma_0}{\mathcal{T}^{1/4}}\,\frac{1}{\tau^{3/4}}\,.
\end{align}
The averaging process uses a large amount of samples at the shorter $\tau$, where the background noise is higher, and progressively smaller amount of samples at longer $\tau$.  
According to \req{eqn:background-m}, it takes $m=10^4$ (2\,H~47\,M) to reduce the background noise by a factor 10, from \hl{$2.1{\times}10^{-14}$} to \hl{$2.1{\times}10^{-15}$ at $\tau=1$} s.

%==================================================
\section{The Experiment}
%==================================================
The block diagram of the experiment follows Fig.~\ref{fig:Experiment-scheme}.
The 100 MHz output of the three CSOs is measured with the TDDS in 6-channel mode, using the covariance.  The 10 GHz outputs are beaten down to HF  and measured with a multi-channel counter.  The latter is a dedicated ``Time and Frequency Monitor'' made by K\&K Messtechnik (now Lange-Electronic \cite{Lange-Electronic}), originally described in \cite{Kramer-2004-EFTF,Kramer-2001-IFCS}.
\hl{The data averaged on $\tau=1$ s are directly available at the output of the the K\&K\@.  With the TDDS, the average on $\tau=1$ s is obtained by decimation of the 10 S/s output stream.} 
We collected all the data measured simultaneously for a duration of $4.7{\times}10^5$ s (5.5 days).

The cutoff frequency $f_H$ is 5 Hz for the TDDS and $0.5$  Hz for the K\,\&,K\@.  This difference is irrelevant because there is no white PM in our results (Sec.~\ref{sec:Results}), thus $f_H$ does not get in the noise equations \cite{Calosso-2016-tuffc}. 

\hl{We had only two synthesizers, thus we used a frquency divider (Hittite) instead.  This divider, at the output of the oscillator $B$, does not have a thermal shield.}

\hl{The oscillator $B$ had the power control not operating properly.  
This problem was discovered when the experiment were already running.}

The most common correlated phenomena, breaking the hypothesis of statistically independent noise processes, are microwave leakage and temperature fluctuations of the environment.
The microwave leakage is in principle absent in our CSOs because the resonator bandwidth is of the order of 10 Hz, a few orders of magnitude smaller than the frequency difference between the oscillators.  
Crosstalk in the instruments is a lesser problem because it impact on the phase, instead of on the frequency.  
Nonetheless, we set the synthesizers at three different frequencies slightly off the nominal value of 100 MHz.  We observed that spurs are also reduced in this way.  
The HF beat notes are substantially immune from leakage, being well separated.

\begin{figure}\centering
\includegraphics[width=88mm]{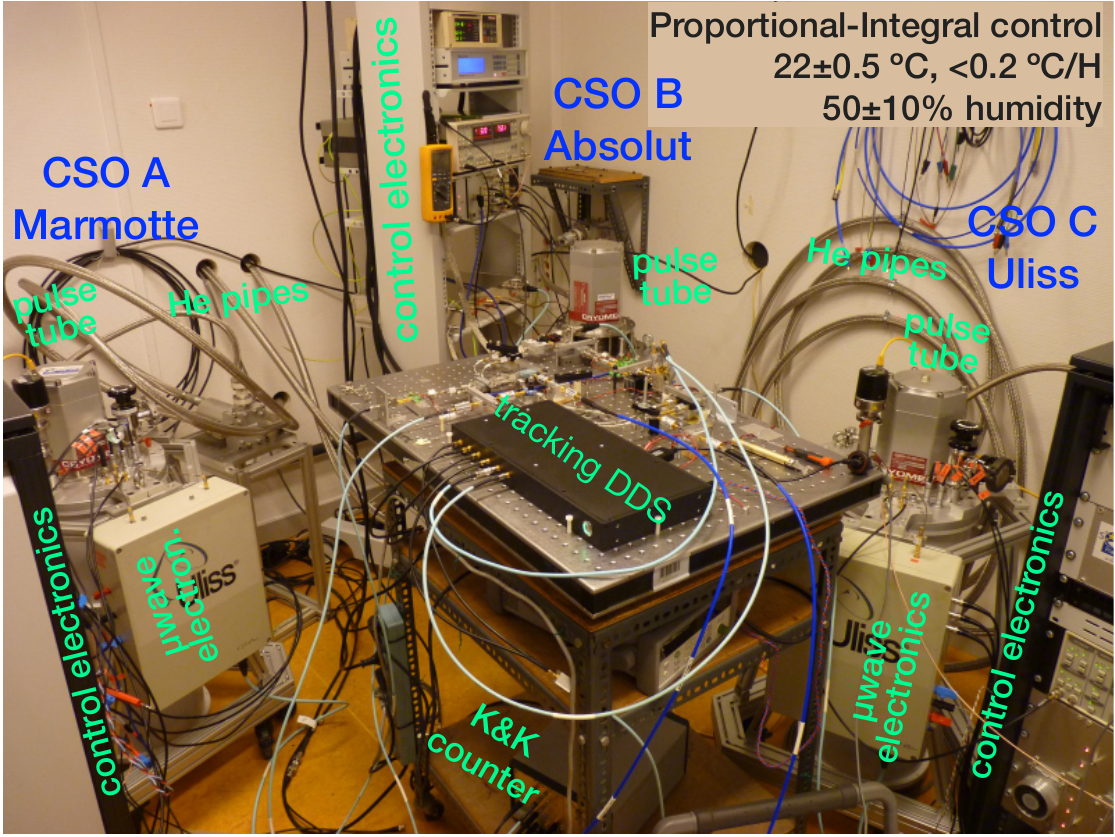}
\caption{The experiment in its environment.  The Helium pumps are in the next room}
\label{fig:Experiment-Photo}
\end{figure}

The experimental setup is shown on Fig.~\ref{fig:Experiment-Photo}.
\hl{The He pumps are located in a nearby room.}
\hl{The thermal fluctuations are strongly reduced by a sophisticated air-conditioning installation.}  
\hl{A proportional integral control guarantees a temperature of $22{\pm}0.5$ \Celsius, with a maximum drift of 0.2 \Celsius/H, and humidity of $50\%{\pm}10\%$.}
\hl{The operators are not present in the room during the measurements.}

%==================================================
\section{Results}\label{sec:Results}
%==================================================
We analyze the results step by step going through consistency checks.  This is necessary because the two-sample variance still being little used, thus we cannot take benefit from the general experience.  

\begin{figure}[t]\centering
\includegraphics[width=88mm]{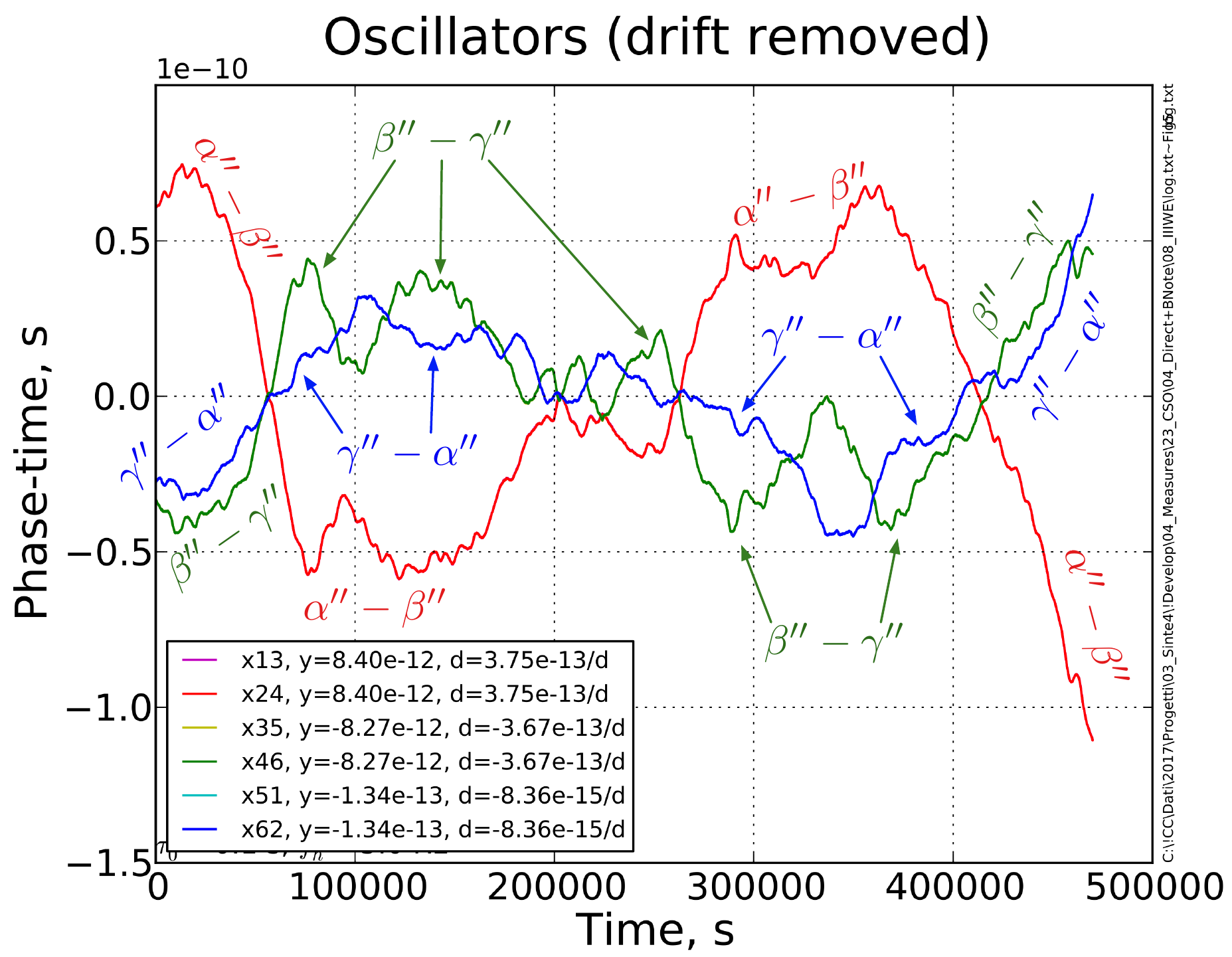}
\vspace*{-2em}
\caption{Time fluctuation of the differences $\color{MidRed}(\sx_{\alpha''}-\sx_{\beta''})\color{black}$, $\color{MidGreen}(\sx_{\beta''}-\sx_{\gamma''})\color{black}$, and $\color{MidBlue}(\sx_{\gamma''}-\sx_{\alpha''})\color{black}$ taken from the TDDS\@.}
\label{fig:x-vs-t}
\end{figure}

The first result (Fig.~\ref{fig:x-vs-t}) is the time fluctuation of the differences $\color{MidRed}(\sx_{\alpha''}-\sx_{\beta''})\color{black}$, $\color{MidGreen}(\sx_{\beta''}-\sx_{\gamma''})\color{black}$, and $\color{MidBlue}(\sx_{\gamma''}-\sx_{\alpha''})\color{black}$ taken from the outputs $\alpha''$, $\beta''$ and $\gamma''$ of the TDDS\@.    
Text colors are consistent with the plots.  At the scale shown, the same differences taken from the outputs $\alpha'$, $\beta'$ and $\gamma'$ overlap, and provide no useful information.  
A visual comparison \hl{between Fig.~{\ref{fig:x-vs-t}} and a set of simulated random time series indicates that the oscillators are mainly affected by white FM and flicker FM}, with some drift showing up.  The peak-to-peak difference is within 100 ps in the first 4 days, and increases slightly afterwards.

The curves are clearly correlated.  By visual inspection on Fig.~\ref{fig:x-vs-t}, the following sums reveal correlation 
\begin{align*}
\color{MidRed}(\sx_{\alpha''}-\sx_{\beta''})\color{black}+
\color{MidGreen}(\sx_{\beta''}-\sx_{\gamma''})\color{black} 
&\approx \text{`small'} && C, A~\text{anticorrelated}\\
\color{MidRed}(\sx_{\alpha''}-\sx_{\beta''})\color{black}+
\color{MidBlue}(\sx_{\gamma''}-\sx_{\alpha''})\color{black}
&\approx \text{`small'} && B, C~\text{correlated}\\
\color{MidGreen}(\sx_{\beta''}-\sx_{\gamma''})\color{black}+
\color{MidBlue}(\sx_{\gamma''}-\sx_{\alpha''})\color{black}
&\approx \text{`large'} && A, B~\text{anticorrelated}\,.
\end{align*}
We believe that this is still a thermal effect on the oscillators, despite the careful conditioning of the room.

\begin{figure}[ht]\centering
\includegraphics[width=83mm]{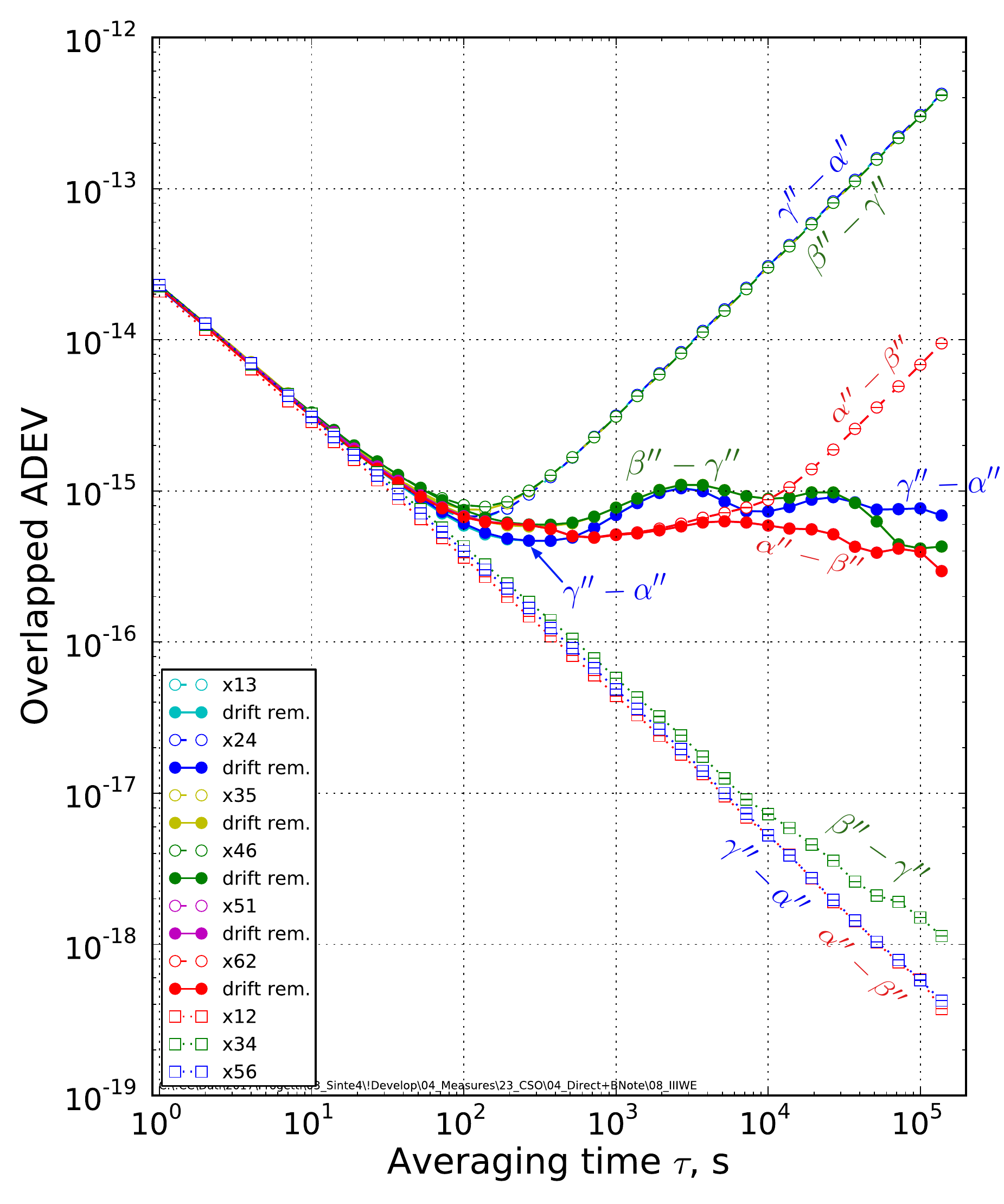}
\vspace*{0em}
\caption{Allan deviation of the difference between oscillators.}
\label{fig:ADEV-Diff}
\end{figure}

The second result (Fig.~\ref{fig:ADEV-Diff}) is the overlapped Allan deviation of some relevant differences, taken at the output of the TDDS\@.  On this figure, we identify three groups of plots.
The first group (the squares, going down to $10^{-18}$\hl{) represent} $\color{MidRed}\sigma_{\alpha'-\alpha''}$, $\color{MidGreen}\sigma_{\beta'-\beta''}$ and $\color{MidBlue}\sigma_{\gamma'-\gamma''}$.  Each curve is the difference between the two channels measuring the same oscillator.  The oscillator fluctuation cancels, and the curve gives the total fluctuation of a pair of channels.  The value \hl{$\sigma_\sy=2.1{\times}10^{-14}$} at $\tau=1$~s is consistent with the flicker PM of the DDS ($\sbb_1=-110$ \unit{dBrad^2}, Fig.~\ref{fig:TDDS-Noise}), within 1 dB\@.
The curves $\color{MidRed}\sigma_{\alpha'-\alpha''}$ and $\color{MidBlue}\sigma_{\gamma'-\gamma''}$ match the prediction based on the dominance of flicker PM, $\sigma^2=\sh_{-1}(3\gamma-\ln(2)+3\ln(2\pi f_H\tau))/(2\pi\tau)^2$.   
The third curve, $\color{MidGreen}\sigma_{\beta'-\beta''}$, gets slightly higher beyond $\tau\approx1000$ s.  
This is due to a defect in channel \hl{$\beta'$}, clearly identified but still not understood.

The second group of plots (the thick dots on Fig.~\ref{fig:ADEV-Diff}) is $\color{MidRed}\sigma_{\alpha''-\beta''}$, $\color{MidGreen}\sigma_{\beta''-\gamma''}$, and $\color{MidBlue}\sigma_{\gamma''-\alpha''}$ of the 100 MHz signal, drift removed.  
These two-sample deviation is dominated by the background noise for $\tau<100$ s.  Beyond, the curves converge to $\color{MidRed}\sigma_{A-B}$, $\color{MidGreen}\sigma_{B-C}$, and $\color{MidBlue}\sigma_{C-A}$, that is, the combined fluctuations of two oscillators each.

The third group of plots (the circles on Fig.~\ref{fig:ADEV-Diff})
is the same as the second, but the drift is not removed.  The drift is clearly visible on the right-hand side of the plot, where the curves are proportional to $\tau$.  Two of these curves, $\color{MidRed}\sigma_{A-B}$ and $\color{MidGreen}\sigma_{B-C}$ reveal a significantly higher drift, due to the oscillator $B$\@.
This is a consequence of the power control not working.
Using $\sigma^2=(1/2)D^2_\sy\tau^2$, the drift of $B$ is $D_\sy=4.5{\times}10^{-18}$/s, or $3.9{\times}10^{-13}$/day.

\begin{figure}[ht]\centering
\includegraphics[width=88mm]{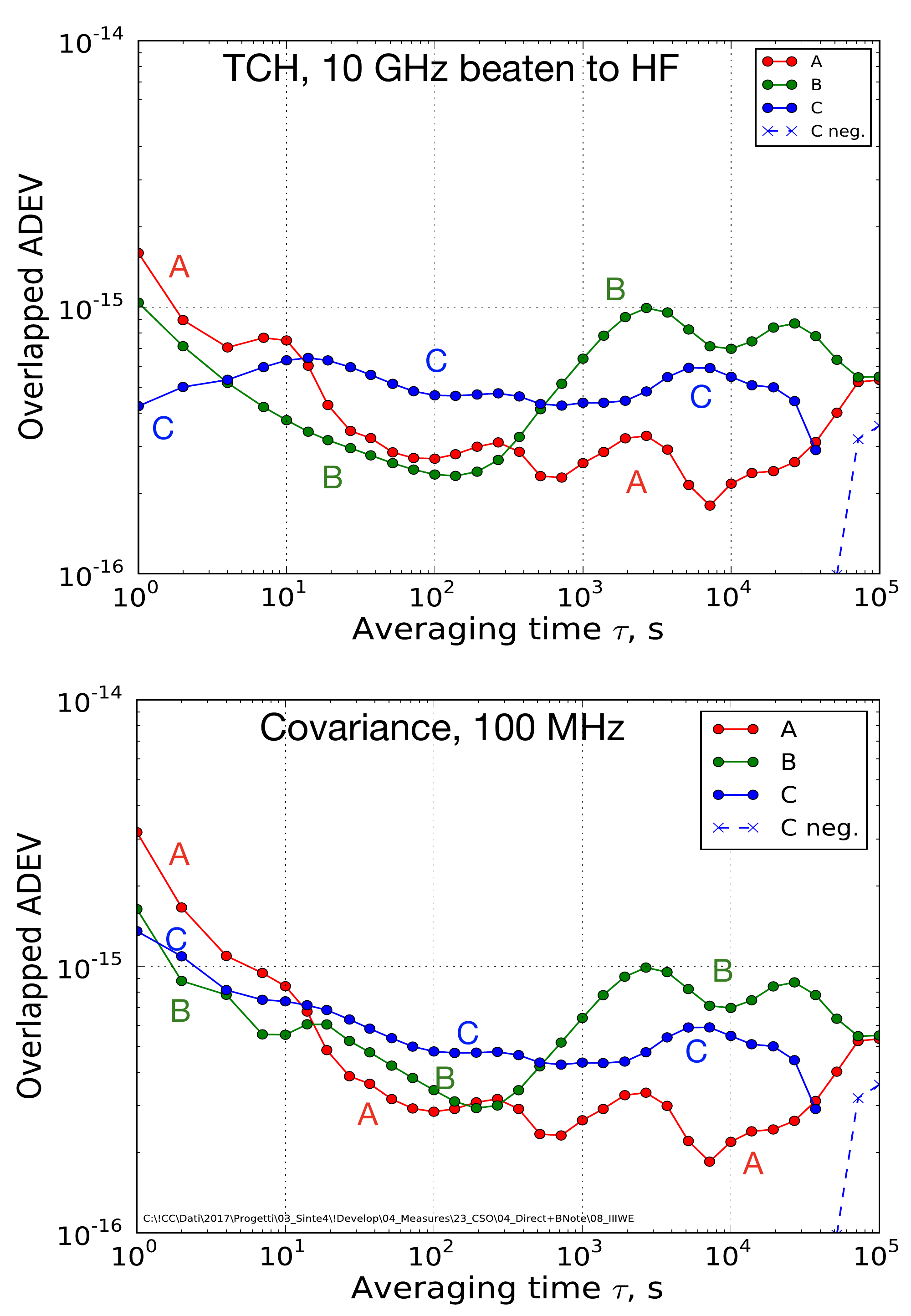}
\caption{Allan deviation of the three oscillators, measured with the Three-Cornered Hat method on the HF beat notes (top), and with the \hl{covariance} method on the 100 MHz outputs (bottom).}
	\label{fig:TCH-GCOV}
\end{figure}

Figure~\ref{fig:TCH-GCOV} shows the ADEV of the three oscillators, measured with the TCH method on the beat notes (top), and (bottom) with the covariance method on the 100 MHz outputs.  The drift is removed in both cases.  
The \hl{results} are substantially equivalent for $\tau>100$ s, while the noise of the synthesizers shows up for shorter $\tau$.
The \hl{Allan deviation plot of the} oscillator $B$ has a bump at $\tau=20$ \hl{on the covariance plot (Fig.~{\ref{fig:TCH-GCOV}} bottom), not present on the TCH plot (Fig.~{\ref{fig:TCH-GCOV}} top).  This is probably due to the frequency divider, which has no thermal shield.} 
\hl{The oscillator $B$ has also irregular behavior and significantly higher instability than the other oscillators} for $\tau>1000$ s.  This is ascribed to the failure in the power control.
The blue line (Oscillator $C$) is dashed for $\tau>5{\times}10^4$ s because a negative covariance appears in the evaluation, and the result makes no sense.  This can be ascribed to the insufficient number of data, or to the presence of slow correlated terms, probably induced by the temperature.

%==================================================
\section{Conclusions}
\begin{figure}[t]\centering
\includegraphics[scale=0.33]{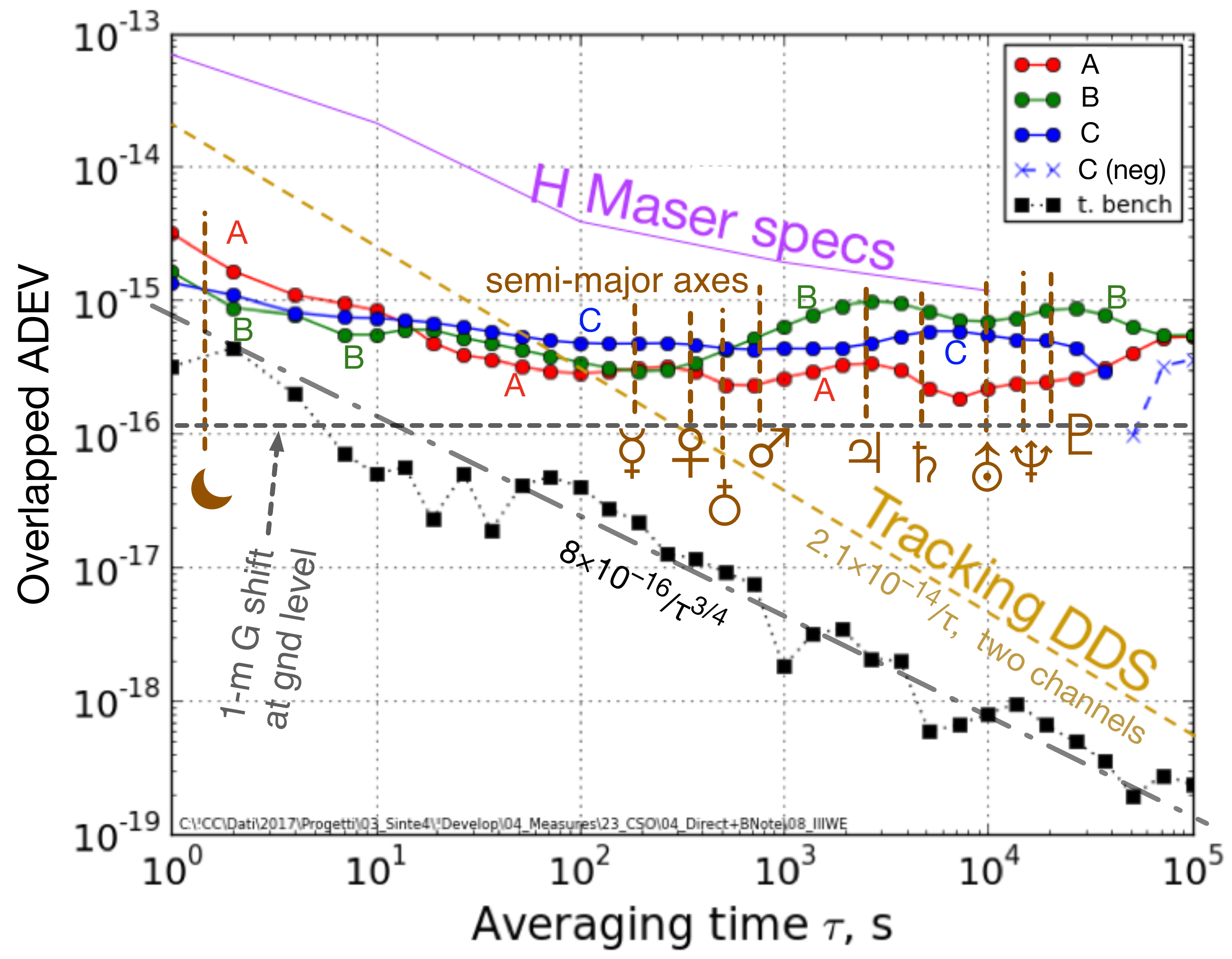}
\vspace*{-2em}
	\caption{The Allan deviation of our oscillators (100 MHz output) compared to some relevant quantities discussed in the text.}
	\label{fig:Results}
\end{figure}

%==================================================
Figure~\ref{fig:Results} summarizes the relevant results obtained with the TDDS and the covariance at the 100 MHz output of the CSOs.

The gold dashed line shows the background noise of the TDDS, accounting for two channels.  This is fair because the simplest measurement, with no correlation, takes two channels.
The background noise is low enough for the measurement of H masers and the other classical atomic standards.  Additionally, the CSOs can be measured in the correlation mode, using two channels per oscillator and the covariance method.

The statistical limit is set by the duration $\mathcal{T}$ of the time series according to \req{eqn:background-T}.  Taking \hl{$\sigma_0=2.1{\times}10^{-14}$}, which is the combined background of two channels, we find \hl{$\sigma=8{\times}10^{-16}/\tau^{3/4}$}.  The theoretical prediction (grey dash-dot line) fits well the observed residuals (dotted line with black squares). 

The RF/microwave hardware can be a serious limitation at this level.  The \hl{phase} fluctuation of the oscillator $C$, $\sigma=1.5{\times}10^{-15}$ at $\tau=1$ s, \hl{corresponds} to a time stability of 1.5 \hl{fs}, or a length stability of 400 nm on a coaxial cable (velocity factor of 0.88 for a 1/2'' Heliax cable).  

The CSO instability is in the low $10^{-15}$ at 1 s, and below $10^{-15}$ for $\tau\ge10$ s up to one day.  For reference, the gravitational shift $g/c^2$ is of $1.09{\times}10^{-16}$ when a clock is raised by 1 m at the ground level.

\hl{The highest stability is seen for $\tau$ between seconds and hours, which is the travel time of the light at interplanetary distances (semi-major axes of the planet orbits on Fig.~{\ref{fig:Results}}). This feature makes the CSO an ideal clock for applications related to the exploration of the solar system.}

%==================================================
\section*{Acknowledgements}
%==================================================
This work is partially funded by (1) the ANR Programme d’Investissement d’Avenir (PIA) under the Oscillator IMP project and the First-TF network, (2) by grants from the Région Bourgogne Franche Comté intended to support the PIA, (3) FEDER, (4) EMRP program (IND55 Mclocks). A special thank to the European Space Agency for supporting this activity.

%==================================================
\bibliography{ref-short,CSO-Meas}
%==================================================

\end{document}